# Infrared Transmissometer to Measure the Thickness of NbN Thin Films

KRISTEN A SUNTER,[1,*] ANDREW E DANE,[1] CHRISTOPHER I LANG,[1] KARL K BERGGREN[1]

[1]*Research Laboratory of Electronics, Massachusetts Institute of Technology, 77 Massachusetts Avenue, Cambridge, MA 02139*
*Corresponding author: k_sunter@mit.edu*

**We present an optical setup that can be used to characterize the thicknesses of thin NbN films to screen samples for fabrication and to better model the performance of the resulting superconducting nanowire single photon detectors. The infrared transmissometer reported here is easy to use, gives results within minutes and is non-destructive. Thus, the thickness measurement can be easily integrated into the workflow of deposition and characterization. Comparison to a similar visible-wavelength transmissometer is provided.**

http://dx.doi.org/x

## 1. INTRODUCTION

Thin films are often used as the starting material for the fabrication of devices such as superconducting nanowire single photon detectors, and it is often necessary to control the thickness of the films to create devices with reproducible performance. Current reports in the literature often control the thickness by relying on the deposition time [1] or do not report how the film thickness was determined and offer only approximations of the thickness based on the deposition time [2,3]. Others have conducted TEM studies on films [4], which are destructive. In our group, we have found that the film thickness does vary consistently with the deposition time, but only as long as the deposition parameters (such as the flow rates of gases) are constant. When the deposition parameters are varied to achieve higher quality films, deposition time cannot be used to compare the thicknesses of films. An independent measurement of the film thickness is also useful to detect drift of the growth system parameters over time, where a constant deposition time may produce different film thicknesses.

Currently, there are several instruments available to determine the thickness (and sometimes simultaneously the refractive index) of thin films using light, but they are more complicated and time-consuming than the optical setups presented here [5-13]. They are based on either ellipsometry or the reflectance and transmittance of samples. Variable angle spectroscopic ellipsometry (VASE) involves scanning a film with a collimated beam over a range of angles of incidence and wavelengths. A model of the film can then be built to fit the resulting reflectance data. The main disadvantage is the cost of a VASE instrument and the amount of time required to scan a single sample, which can be on the order of hours depending on how detailed the analysis has to be and how much is known about the film *a priori*. Thin films used in VASE must also be on substrates that are not transparent to the wavelengths scanned, which is a problem for work with photodetectors.

Some commercial reflectometers measure the reflectance and transmittance of a sample at different wavelengths and include software to analyze the results to determine the index of refraction and thickness for some material systems down to 1 nm. To our knowledge, they have not been used successfully to develop a niobium nitride deposition process. However, the disadvantage of a commercial system is the cost, especially when the instrument will only be used for a limited range of samples, and advanced software for determining the composition of the thin film stack is necessary.

Other researchers have reported optical setups that can determine the thickness and refractive index of films in a non-destructive manner, but these setups are generally more complicated than ours due to the specific problems investigated. For example, Hirth et al. [5] combine reflectometry and confocal microscopy to determine both film thickness and topography. Jafarfard et al. [6] use dual-wavelength diffraction phase microscopy to determine the refractive index and the thickness spatial distribution of a sample, which requires the use of a laser, a transmission grating, a spatial filter and collimating and focusing optics. Joo et al. [7] consider angle-resolved reflectometry at different wavelengths, similar to a VASE, and Henrie et al. [8] created a spectral reflectometer with a series of LEDs to cover the visible wavelength region. Others have built far ultraviolet (FUV) [9] or extreme ultraviolet (EUV) [7,10-13] reflectometers, which require a laser or other source that operates at a wavelength of tens of nanometers and in some cases controlling the angle of incidence. None of these approaches have the simplicity and accuracy of a basic transmissometer for the problem of determining absorbing film thicknesses in the few nanometer range.

## 2. TRANSMISSOMETER SETUP

### A. Infrared Transmissometer

The concept behind the operation of the transmissometers is shown in Figure 1. Quantifying the amount of transmitted light should lead to a relative measure of thickness between films of the same material. Light from the LED source is incident on the sample: this figure shows a non-normal angle of incidence for clarity, but, in our case, the transmissometer used normal incidence. The light that is transmitted through the sample is detected to determine the transmittance ($T$) of the film. As shown in the schematic, light is both transmitted and reflected at the front and back of the sample, and multiple passes of light contribute to the overall transmittance. Some of the light is also absorbed in the NbN layer.

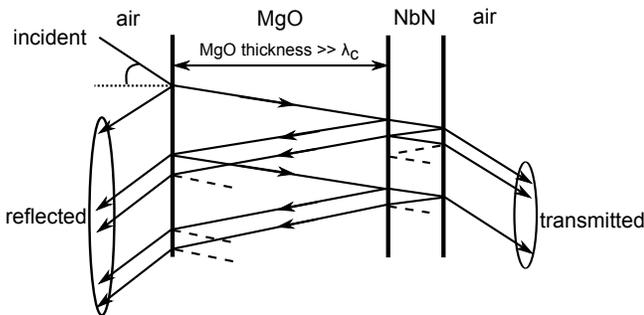

Fig. 1. Illustration of how the internal reflections within a substrate contribute to the total reflectance and transmittance of a sample measured in an optical setup. The optical setups described here rely on normal incidence ($\theta=0°$).

The total amounts of light transmitted, reflected and absorbed by a sample can be calculated analytically. The analytical optical model uses the Fresnel equations to model the reflectance and transmittance of light at the interface between the air and the substrate and the transfer matrix method to model the transmittance, reflectance and absorptance of the thin film on the other side of the substrate [14]. The analytical optical model requires the index of refraction of each material in the sample. These values can either be found in the literature or measured for a typical sample using VASE. In our case, we relied on literature values for the index of refraction of the substrate materials (silicon, silicon dioxide and sapphire), and the complex index of refraction of NbN was based on the value for a single thick NbN film measured by J.A. Woollam Co. Once the calculated transmittance versus NbN film thickness was plotted for a particular substrate, the transmittance of our samples was measured, and the thicknesses were read from the transmittance versus film thickness plot, as shown in Figure 2(a).

Figure 2(b) shows the basic transmissometer setup for an infrared (1550 nm) transmissometer built in our laboratory. This wavelength was selected because silicon is transparent in the infrared, and many of our devices are grown on silicon substrates. In addition, our devices are tested at 1550 nm because it is an important wavelength for telecommunication, which is one of the applications of SNSPDs. Even if there is some error in the optical constants of our NbN thin films because the density of thin films is different than that of the thicker film used to determine the index of refraction of NbN, the thickness calculated by the transmissometer should reflect an ``optical thickness'' that might not match the physical thickness but should lead to the correct calculated absorptance in the optical models of device performance. However, there may still be a difference between the optical constants at room temperature, where the films are measured, and at low temperature, where devices are operated.

The transmissometer consists of a single column with an LED, a lens to collimate the LED light, a sample stage and a lens to focus the LED light onto the detector. An LED is used instead of a laser because the coherence length of the LED light is smaller than the thickness of the substrate, and thus there are no etalon effects within the substrate and its thickness does not need to be known.

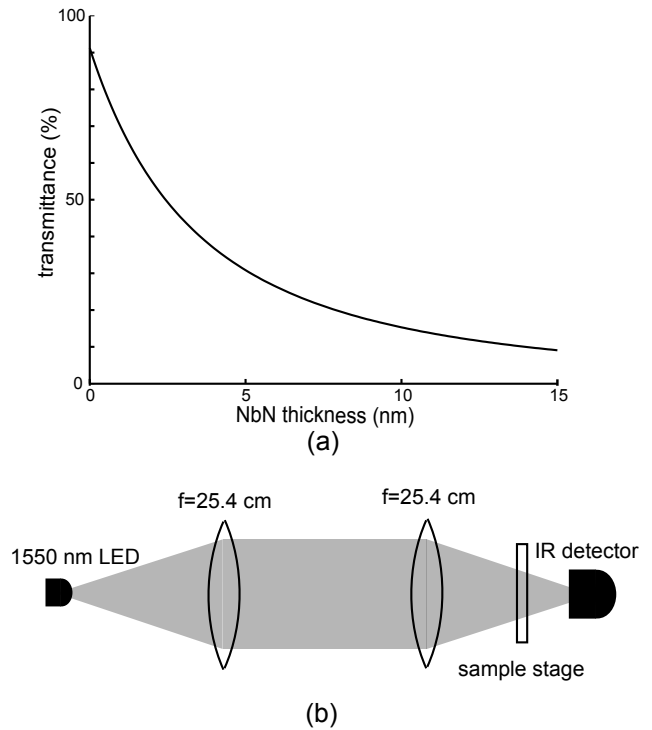

Fig. 2. (a) Calculated transmittance versus NbN thickness for an NbN film grown on 225-nm-thick silicon oxide on silicon, with 225-nm-thick silicon oxide on the back of the substrate. (b) Schematic of the infrared transmissometer showing the basic optical components.

To perform measurements, first the dark signal from the detector, taken when the LED is off, was recorded. Then, the detector was read (1) when there was no sample mounted, (2) when a blank substrate was mounted, and (3) when the chip under test was mounted. The chip was typically mounted with the film side down, facing the detector, to prevent the accumulation of dust on the film surface. These measurements were then repeated for different LED powers. The dark signal was subtracted from all recorded data. Then, the measured signal with the blank substrate mounted divided by the signal with no substrate mounted gave the transmittance through the blank substrate. This value was used to determine whether the optical constants of the substrate used in the calculation were accurate. For a double-sided polished silicon substrate composed of 225 nm of silicon nitride on both sides, the transmittance should be approximately 56.4%. The transmittance was determined by the detector signal with the NbN sample mounted divided by the signal with no sample, and it was used to determine the thickness of the NbN, given the plot of the transmittance versus NbN thickness.

### B. Visible Transmissometer

A second instrument operating at 470 nm was built to compare the results measured at different wavelengths. One advantage of operating at 470 nm is that visible-wavelength photodetectors are less expensive than IR photodetectors. However, 470-nm light cannot be used to measure films grown on silicon substrates because silicon is not transparent in the visible, so the IR transmissometer is necessary for silicon substrates. Another disadvantage of the visible transmissometer is that room lights lead to noise at 470 nm but not at 1550 nm, so the visible transmissometer required shielding and the IR transmissometer did not. The two setups were expected to be sensitive to differences in NbN thickness at different thickness ranges. As shown in Figure 3, the slope of the transmittance versus thickness of NbN on

MgO is steeper at 1550 nm for small thicknesses, which implies that a higher precision is possible. However, for larger NbN thicknesses, the transmittance at 1550 nm changes more slowly with thickness and is lower in magnitude than that at 470 nm, and thus the visible transmissometer can more accurately determine the thickness of films thicker than 10 nm. Thus, each transmissometer is suited for different types of samples.

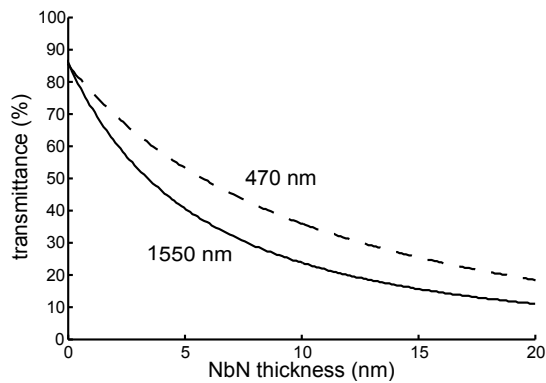

Fig. 3. Calculated transmittance versus NbN thickness of an NbN film on MgO for 470-nm light (circles) and for 1550-nm light (diamonds). The steeper slope of the 1550-nm line at small thicknesses suggests that it should be more sensitive to differences in thickness of very thin films.

Figure 3 relies on a model that does not include an anti-reflection coating (ARC), which would typically be spun onto the substrate at the end of the fabrication process and not immediately after film growth, when the films are characterized. The model also neglects an oxide layer on the NbN. Niobium does oxidize in air, though the samples are characterized soon after growth and stored in a nitrogen box to limit oxidation. In addition, the oxide layer, which can be as much as 2 nm thick according to previous TEM results [15], does not significantly affect the transmission characteristics. For example, the transmittance of a 4.00-nm-thick layer on MgO is 58.5% (illuminated through the substrate), and adding a 2.00-nm-thick layer of niobium oxide only increases the transmittance to 58.9%. This transmittance value would correspond to a NbN thickness of 3.92 nm in the model without the niobium oxide, which is a decrease in measured thickness of less than 0.100 nm. Therefore, the niobium oxide layer is neglected in the plots used here to determine the NbN thickness.

## 3. EXPERIMENTAL RESULTS

The experimental results using the transmissometer show that the infrared and optical transmissometers give repeatable values for the thickness and offer a quick, non-destructive method to ascertain the relative thicknesses of films. The measurements using the visible reflectometer and IR transmissometer were compared to the sheet resistance of films, the deposition time and VASE measurements.

### A. Film Thickness Versus Sheet Resistance

The sheet resistance should vary inversely with the thickness of the films, and the correlation between the sheet resistance and a measurement of the thickness should therefore be high. Figure 4 shows the sheet resistance as measured with a four-point probe setup versus the deposition time and the sheet resistance versus film thickness determined with the transmittance measured by the visible-wavelength transmissometer. The inverse of the sheet resistance correlates much better with the thickness than the deposition time, with $R^2$ values of the linear fit of 0.65 and 0.06, respectively.

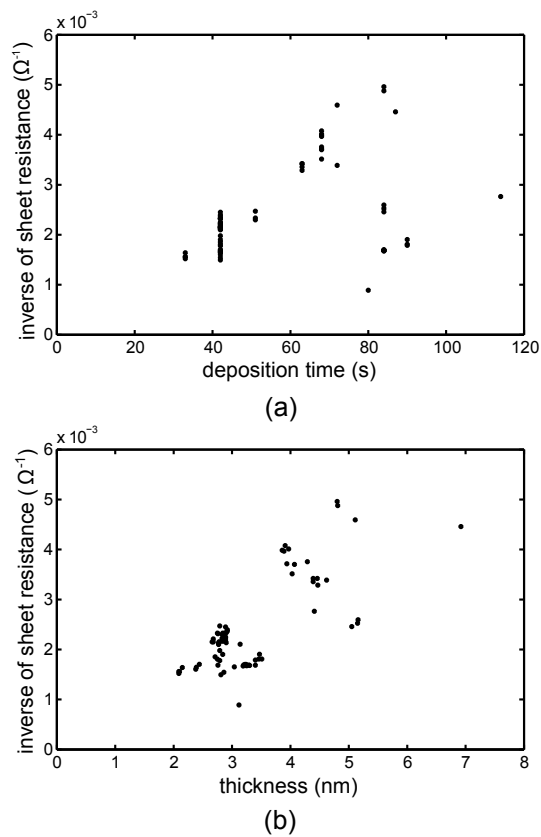

Fig. 4. (a) Sheet resistance versus deposition time. A linear fit to the data gave an $R^2$ value of 0.1008. (b) Sheet resistance versus film thickness determined from transmittance measurements in the visible (470-nm) transmissometer. A linear fit to the data yielded an $R^2$ value of 0.5483.

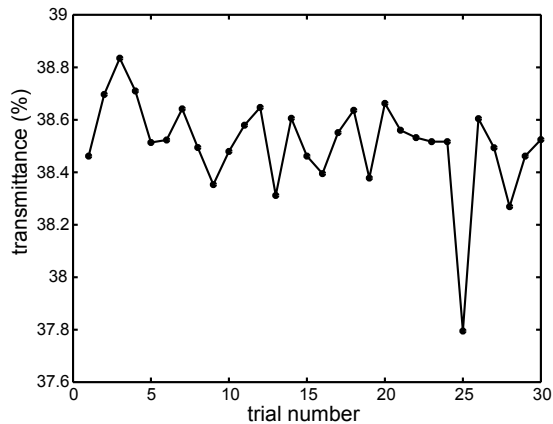

Fig. 5. Transmittance values measured with the IR transmissometer of an NbN film on MgO. Thirty measurements were performed in succession to demonstrate the repeatability of the measurements.

### B. Repeatability of Thickness Measurements

Figure 5 shows the results of thirty measurements of a film on MgO taken in succession to demonstrate the repeatability of the measurements. For each measurement, the detector reading with no sample mounted was taken first. Then, the sample was mounted, the detector reading was taken again, and the sample was unmounted before beginning the next measurement. The results of this ironman trial show that the measurements are repeatable, without short-term drift. The thickness values for these transmittance values range from

4.875 nm to 5.075 nm, and so the thirty trials are within 0.2 nm of each other.

Figure 6 shows the film thicknesses of eight NbN films on MgO as measured with the IR transmissometer. The second set of measurements was taken four weeks after the first set to demonstrate that the IR transmissometer gives repeatable results over time. Each of the films appears to have decreased in thickness, which is likely due to oxidation over time. The greatest variation was still less than 0.2 nm.

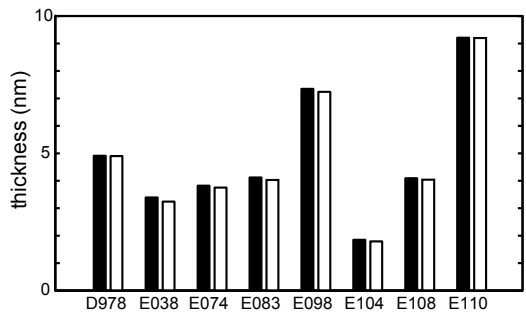

Fig. 6. Thicknesses of several NbN films grown on MgO. The first set of measurements is given in black, and the second set, 28 days later, is given in white. The films are slightly thinner in the later measurements, likely indicating oxidation over time.

### C. Comparison of the Results of the Different Optical Setups

The thickness of films were measured in the visible and infrared optical setups to determine whether the setups gave equally consistent results, and the results are shown in Figure 7. Figure 7(a) shows the film thicknesses versus deposition time for different substrates measured in both the visible and infrared setups. The IR transmissometer results were consistently thicker than the visible transmissometry results for NbN on MgO, but all films show the expected linear increase in thickness with deposition time. In addition, the thicknesses of NbN films on different substrates are consistent for similar deposition times. It is interesting that the thicknesses of NbN on MgO as measured in the IR transmissometer were consistently thicker than NbN films on silicon for the same deposition time. This difference could be due to an uncertainty in the optical constants of the materials, but it might also be due to a different microstructure in the NbN when grown on a crystalline substrate (such as MgO) compared to films grown on amorphous substrates (such as silicon nitride or silicon oxide on silicon), which might lead to either a different thickness or different optical constants of the NbN.

Figure 7(b) shows the thickness data points for the samples of NbN on MgO as determined in the IR transmissometer and the visible reflectometer. A linear fit of the data suggests that the thickness measurements are consistent between the two optical setups, although one gives consistently larger film thicknesses. The deviation of the thickness measurements between the optical setups is likely due to the uncertainty in the index of refraction of NbN thin films at either wavelength, but the close comparison between the setups indicates that either can be used for relative thickness measurements.

### D. VASE Results

A variable angle spectroscopic ellipsometer (VASE) was used to characterize a thin film of NbN on silicon nitride on silicon, which could give both the thickness and the optical constants of NbN in the visible. The optical constants in the infrared could not be determined with VASE because silicon is transparent in the infrared, so the VASE results were only compared to the results obtained with the visible-wavelength transmissometer. A MgO chip that was sputtered in the same deposition run as the silicon nitride on silicon chip measured with the VASE was found to have a film thickness of 4.6 nm according to the visible transmissometer, which compares to a thickness of approximately 8 nm found for the film on silicon nitride with the VASE. One cause of this discrepancy was probably due to the difference between the optical constants found for NbN using the VASE and those used to calculate the thickness according to the transmittance found by the reflectometer. This discrepancy is discussed further below.

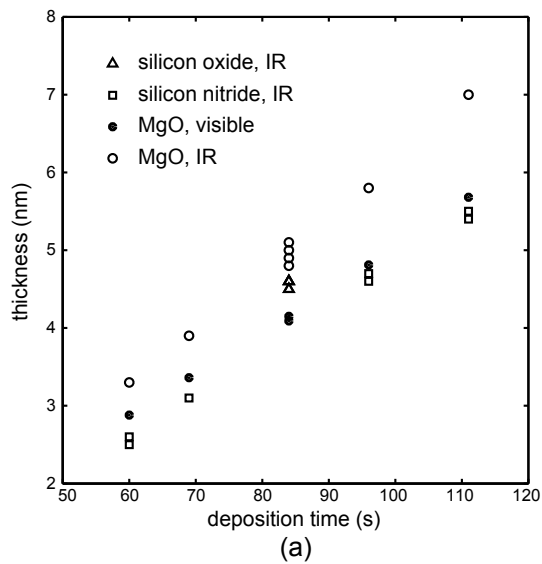

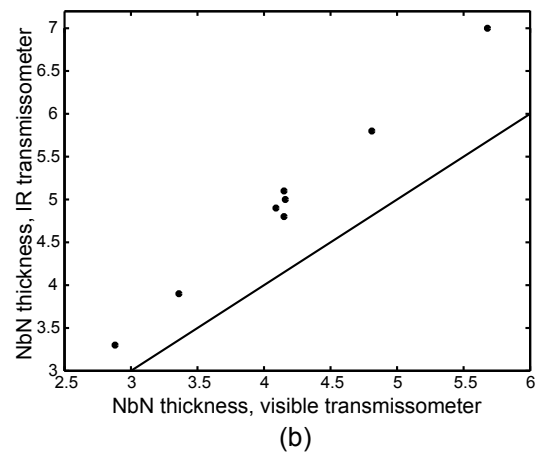

Fig. 7. (a) Film thickness versus deposition time for different substrates: silicon oxide/silicon (triangles), silicon nitride/silicon (squares) and MgO (filled circles) measured with the IR transmissometer, and MgO (open circles) measured with the visible transmissometer. (b) Thickness according to the IR transmissometer versus thickness according to the visible reflectometer for NbN films grown on MgO. The black line is the equality ($y=x$) line. The primary source of uncertainty in the measuremtns is the index of refraction. An additional error is due to the reproducibility of the tool, which was found to be 0.2 nm in the ironman trial.

First, two bare silicon nitride films on silicon were characterized with VASE. The silicon nitride layer was modeled as a Cauchy layer, and the silicon was modeled using parameters in the literature [16]. Although both films were grown using the same CVD system, the Cauchy parameters were not identical. Next, an NbN film grown on a silicon nitride film on silicon was measured. The resulting data were fit using the results for the bare silicon nitride film (using a thickness of 388 nm) and a point-by-point fit to the data. Then, a Drude model fit for the NbN layer was performed, which started with the values for $n$ and $k$ found by the point-by-point fit. The point-by-point fit and the Drude model fit gave NbN thicknesses of 8.12 nm and 8.08 nm, respectively. The mean squared error of the point-by-point fit was 0.767 nm$^2$, and that of the Drude model was 3.46 nm$^2$. As mentioned above, a sister chip of MgO had a deposited NbN thickness of 4.6 nm according to the visible reflectometer, which is over 3 nm thinner. This difference is large, but the cause is not clear. It is possible that the index of refraction used to calculate the thickness according to the visible reflectometer is not accurate. It is also possible that films grown on amorphous substrates, such as silicon nitride, have a different crystalline structure than those grown on nearly lattice matched substrates such as MgO, leading to either a different thickness or different optical constants. Techniques to address these possibilities are discussed below.

## 4. DISCUSSION

The precision of the results according to the optical setups described here is within 0.2 nm, which is the difference between the minimum and the maximum thicknesses over 30 measurements according to the ironman results. However, the accuracy of our optical measurements is unknown because of the uncertainty in the optical constants. In addition, two improvements to the setup are discussed below, involving the spot size and the angle of incidence.

Figure 8 shows the transmittance versus NbN thickness curves for the visible reflectometer for various optical constants of NbN estimated from figures in the dissertation of M. Benkahoul [17] for three different NbN phases: the hexagonal δ'-NbN phase, the cubic δ-NbN phase, which has a NaCl structure composed of FCC sublattices of niobium and nitrogen, and the hexagonal β-$Nb_2N$ phase [18]. Figure 8 also gives the expected results using the values found for films grown by our group [15]. The variation illustrated in this figure is likely similar to what would be found for optical constants measured at other wavelengths as well. It is encouraging that the results using the indexes of refraction measured for different films grown by our group are similar.

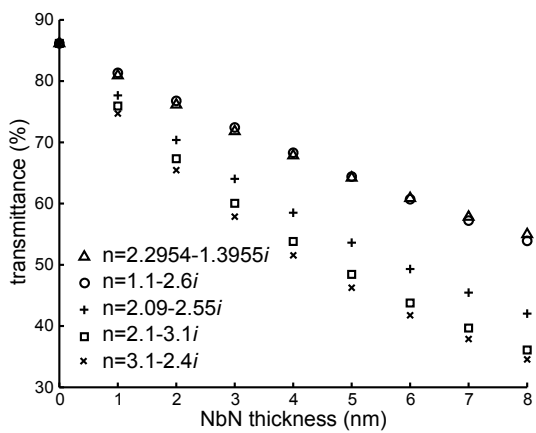

Fig. 8. Calculated transmittance for different optical constants of NbN]{Transmittance versus thickness of NbN on MgO for different optical constants of NbN at 470 nm, showing how much the transmittance and therefore the thickness measured by the transmissometer can vary for different optical constants found in the literature: δ'-NbN: $n$=3.1-2.4$i$ (×), δ-NbN: $n$=2.1-3.1$i$ (□), β-$Nb_2N$: $n$=1.1-2.6$i$ (o). This plot also includes values found for films in our group: $n$=2.09-2.55$i$ (+), $n$=2.2954-1.3955$i$ (Δ).

An MgO chip was deposited with NbN at the same time as the silicon nitride/silicon chip characterized by VASE. The NbN on the MgO chip was measured to be 4.6 nm thick according to the visible wavelength (470 nm) transmissometer when the index of refraction of NbN was set to $n$=2.09-2.55$i$. This value was determined by VASE on a NbN film that was thick enough to be opaque, but the optical constants of thin and thick films can differ considerably [19]. If an index of refraction of $n$=1.9533-1.6933$i$ is used in the mathematical model instead, which was the value found for the thin NbN film on silicon nitride/silicon by VASE, the calculated thickness of the NbN on the MgO sister chip increases to 7.5 nm, which is only 0.6 nm less than the thickness on the silicon nitride/silicon chip determined with VASE. Therefore, the choice of optical constants in the calculation of the transmittance versus thickness is very important for the accuracy of the measurement. However, even with incorrect or unknown optical constants, the transmissometer measurements give a useful indication of the relative thickness of samples. The optical setups can thus be used to provide feedback on the stability of the film deposition process and can be correlated with high-efficiency devices to select the best deposition conditions.

In addition, thin films often have somewhat different structure than the surface of thick films or bulk samples due to the growth of columnar grains and other mesoscale structures [20,21], which could lead to variations in the optical constants themselves with film thickness. There might also be a difference between the microstructures, and thus optical constants, of NbN films grown on crystalline substrates (*e.g.*, MgO or sapphire) and those grown on amorphous substrates (silicon nitride or silicon oxide on silicon). A further study to characterize the optical constants of NbN films of varying thicknesses on different substrates using the VASE could give a more accurate value for the optical constants to be used in the model.

The thickness as measured by the optical setups clearly correlates better with the sheet resistance than the deposition time does, which makes sense given the amount of control over the deposition parameters affecting each individual chip in a run. The TEM comparisons also corroborate the thickness measurements. The VASE results seem the most accurate for the individual film tested, though the method has several drawbacks, one of which is the inability to take data at wavelengths at which the substrate is transparent, such as 1550 nm. Thus, several films from different runs would have to be tested to see whether the relative thicknesses as measured by VASE matched the relative thicknesses measured using the optical setups, perhaps leading to a reliable correlation factor that could be used to calibrate the thickness measurements of the optical setups.

## 5. CONCLUSION

We presented an optical setup to non-destructively measure the film thickness of NbN thin films. The optical measurements correlate better with the sheet resistance than the deposition time, which implies that the deposition time is not as reliable an indicator of the film thickness as the results of the optical measurements. The optical measurements are also simple and relatively quick to perform. However, several assumptions in the optical model used to correlate the measured transmittance with a film thickness lead to discrepancies in the thicknesses measured by two optical setups at different wavelengths and by different methods such as VASE. The model assumes that the optical constants of NbN are constant throughout the film, which might not be the case if the microstructure of NbN changes near the substrate, and future work can investigate models that include variation in the index of films with thickness. In addition, the optical constants of our NbN thin films are not known, and the literature and VASE measurements on other films suggest that they might vary significantly depending on the phase of NbN and the deposition parameters. Despite these issues, the optical results are consistent over time and give a good measure of the relative thickness of films, which is adequate for device development. In the future, the use of optical constants from films measured with VASE can increase the accuracy of the thickness measurement method presented if needed.

## Funding Information

Intelligence Advanced Research Projects Activity/Air Force Research Laboratory (IARPA/AFRL) (FA8650-11-C-7105)

## Acknowledgments

This project was supported by the Intelligence Advanced Research Projects Activity (IARPA) via Air Force Research Laboratory (AFRL) contract number FA8650-11-C-7105. The U.S. Government is authorized to reproduce and distribute reprints for Governmental purposes notwithstanding any copyright annotation thereon. The views and conclusions contained herein are those of the authors and should not be interpreted as necessarily representing the official policies or endorsements, either expressed or implied, of IARPA, AFRL, or the U.S. Government.

Tim Tiwald at JA Woollam Co. performed the fitting to the optical models for the VASE results; JA Woollam also characterized a thick film grown by our group, as did Dr. Gale Petrich at MIT. Chris Lang helped to build both the visible and infrared transmittometers. We also thank Melissa Hunt for building another version of the transmissometer.

## References

1. X. Q. Jia, L. Kang, M. Gu, X. Z. Yang, C. Chen, X. C. Tu, B. B. Jin, W. W. Xu, J. Chen, and P. H. Wu, "Fabrication of a strain-induced high performance NbN ultrathin film by a $Nb_5N_6$ buffer layer on Si substrate," Semiconductor Science and Technology **27**, 035010 (2014)
2. L. Zhang, Q. Zhao, Y. Zhong, J. Chen, C. Cao, W. Xu, L. Kang, P. Wu, and W. Shi, "Single photon detectors based on superconducting nanowires over large active areas," Applied Physics B **97**, 187-191 (2009)
3. M. Hofherr, D. Rall, K. Il'in, A. Semenov H.-W. Huebers, and M. Siegel, "Dark count suppression in superconducting nanowire single photon detectors," Journal of Low Temperature Physics **167**, 822-826 (2012)
4. Y. Ufuktepe, A. H. Farha, S. I. Kimura, T. Hajiri, K. Imura, M. A. Mamun, F. Karadag, A. A. Elmustafa, and H. E. Elsayad-Ali, "Superconducting niobium nitride thin films by reactive pulsed laser deposition," Thin Solid Films **545**, 601-607 (2013)
5. F. Hirth, T. C. Buck, A. P. Grassi, and A. W. Koch, "Depth-sensitive thin film reflectometer," Measurement Science and Technology **21**, 125301 (2010)
6. M. R. Jafarfard, S. Moon, B. Tayebi, and D. Y. Kim, "Dual-wavelength diffraction phase microscopy for simultaneous measurement of refractive index and thickness," Opt. Lett. **39**, 2908-2911 (2014)
7. W.-D. Joo, J. You, and Y.-S. Ghim, "Angle-resolved reflectometer for thickness measurement of multi-layered thin-film structures," Proc. SPIE **7063**, 70630Q (2008)
8. J. Henrie, E. Parsons, A. R. Hawkins, and S. M. Schultz, "Spectrum sampling reflectometer," Surface and Interface Analysis **37**, 568-572 (2005)
9. J. A. Aznarez, J. I. Larruquert, and J. A. Mendez, "Far-ultraviolet absolute reflectometer for optical constant determination of ultrahigh vacuum prepared thin films," Review of Scientific Instruments **67**, 497-502 (1996)
10. S. Doring, F. Hertlein, A. Bayer, and K. Mann, "EUV reflectometry for thickness and density determination of thin film coatings," Applied Physics A **107**, 795-800 (2012)
11. FORMAT FIXG. H. Ho, F.-H. Kang, and H.-W. Fu, "Absorption and loss of film thickness in photoresists and underlayer materials upon irradiation at 13.5 nm," Proc. SPIE **7636**, 76362U (2010)
12. M. Banyay and L. Juschkin, "Table-top reflectometer in the extreme ultraviolet for surface sensitive analysis," Applied Physics Letters **94**, 063507 (2009)
13. S. Schroeder, T. Feigl, and A. Duparre, "EUV reflectance and scattering of Mo/Si multilayers on differently polished substrates," Optics Express 15, 13997-14012 (2007)
14. A. Yariv and P. Yeh, *Optical Waves in Layered Media* (Wiley, 1988)
15. V. Anant, A. J. Kerman, E. A. Dauler, J. K. W. Yang, K. M. Rosfjord, and K. K. Berggren, "Optical properties of superconducting nanowire single-photon detectors," Optics Express **16**, 10750-10761 (2008)
16. C. M. Herzinger, B. Johs, W. A. McGahan, J. A. Woollam, and W. Paulson, "Ellipsometric determination of optical constants for silicon and thermally grown silicon dioxide via a multi-sample, multi-wavelength, multi-angle investigation," Journal of Applied Physics **83**, 3323-3336 (1998)
17. M. Benkahoul, *Niobium Nitride Based Thin Films Deposited By DC Reactive Magnetron Sputtering: NbN, NbSiN and NbAlN* (Ph.D. thesis, Ecole Polytechnique Federale de Lausanne, 2005)
18. M. Torche, G. Schmerber, M. Guemmaz, A. Mosser, and J. C. Parlebas, "Non-stoichiometric niobium nitrides: structure and properties," Thin Solid Films **436**, 208-212 (2003)
19. O. S. Heavens, *Optical Properties of Thin Solid Films* (Courier Dover Publications, 1991)
20. J. A. Thornton, "High rate thick film growth," Annual Review of Materials Science **7**, 239-260 (1977)
21. R. Messier, A. P. Giri, and R. A. Roy, "Revised structure zone model for thin film physical structure," Journal of Vacuum Science and Technology **2**, 500-503 (1984)